\documentclass{iaus}
\usepackage{graphicx}
\newcommand{\cmtwo}{cm$^{-2}$}  
\newcommand{\cmthree}{cm$^{-3}$}

\newcommand{\kms}{km\,s$^{-1}$}       
\newcommand{\ta}{$T_{\rm A}$}        

\newcommand{\tr}{$T_{\rm R}$}

\newcommand{\um}{$\mu$m}                                 
\newcommand{\molh}{H$_{2}$}                              

\newcommand{\water}{H$_{2}$O}

\newcommand{\molo}{{\rm O}$_2$}
\newcommand{\otwofull}{${\rm O}_2\,(N_J=1_1-1_0)$}
\newcommand{\roc}{$\rho \, {\rm Oph \, cloud}$}
\newcommand{\roa}{$\rho \, {\rm Oph \, A}$}
\newcommand{\roac}{$\rho \, {\rm Oph \, A \, cloud}$}

\newcommand{\amin}{$^{\prime}$}                   
\newcommand{\asec}{$^{\prime \prime}$}

\newcommand{\amindot}[2]{\mbox{#1$\stackrel {\prime}{_{\bf \cdot}}$#2}}

\title[O$_2$ in $\rho$ Oph A] 
{Odin \thanks{Odin is a Swedish-led satellite project funded jointly by 
  the Swedish National Space Board (SNSB), the Canadian Space Agency (CSA), 
  the National Technology Agency of Finland (Tekes) and Centre National
  d'Etude Spatiale (CNES).} detection of O$_2$}

\author[R. Liseau et al.]   
{R. Liseau   
\and the O$_2$din Team}

\affiliation{Stockholm Observatory, AlbaNova University Center, Stockholm University, Sweden \break email: rene@astro.su.se
}

\pubyear{2005}
\volume{231}  
\pagerange{1--1}
\date{?? and in revised form ??}
\jname{Proceedings Title IAU Symposium}
\editors{A.C. Editor, B.D. Editor \& C.E. Editor, eds.}
\begin{document}

\maketitle

\begin{abstract}
We present the detection of molecular oxygen with Odin toward the dense molecular core \roa, which is part of a region of active star formation. The observed spectral line is the $(N_J=1_1-1_0)$ ground state transition of \molo\ at 119\,GHz (2.5\,mm wavelength). The center of the line is at the LSR velocity of a number of optically thin lines from other species in the region and the \molo\ line also has a very similar, narrow, line width. Within the 10\amin\ beam, the line intensity is $\int\!T_{\rm A}\,{\rm d}v = 28$\,mK \kms, which corresponds to $5\,\sigma$ of the rms noise. A standard LTE analysis results in an \molo\ abundance of $5 \times 10^{-8}$, with an uncertainty of at least a factor of two. We show that standard methods, however, do not apply in this case, as the coupling of the Odin beam to the source structure needs to be accounted for. Preliminary model results indicate \molo\ abundances to be higher by one order of magnitude than suggested by the standard case. This model predicts the 487\,GHz line of \molo\ to be easily detectable by the future Herschel-HIFI facility, but to be out of reach for observations on a shorter time scale with the Odin space observatory.
\keywords{ISM: molecules - lines and bands - abundances - ISM: clouds - stars: formation}
\end{abstract}

\noindent
{\bf subject index:} oxygene reservoirs - molecular processes - molecular excitation - dense clouds - star formation \\
{\bf molecular index:} \molo \\
{\bf object index:} \roa
 
\section{Introduction}

The energy balance of the interstellar medium (ISM) relies on a complex interaction of various heating and cooling processes, which of course also affect the chemistry being activated in the medium.This in turn regulates the amount of the major cooling species in the gas. Heating procsses include mechanical and radiativ energy input, whereas, in dense clouds, the cooling is dominated by molecules and dust particles, resulting in low equilibrium temperatures on short time scales. Whereas the coolant CO is readily observable from the ground, water and oxygen molecules need to be observed from space. 

Odin is thus a mission in orbit around the Earth dedicated for the observation of \water\ and \molo\ and the facility is shared by aeronomers and astronomers on an equal time basis. The millimeter/submillimeter (mm/submm) observatory Odin is described by Frisk et al. (2003) and Olberg et al. (2003), whereas Odin-astronomy is reviewed by Hjalmarsson et al. (2003). These articles are all collected in the special Odin edition of Astronomy \& Astrophysics, which of course also contains numerous papers on initial results obtained with Odin. Energy level and excitation diagrams for the principal observable molecules can be found in the review by Liseau (2001).

\begin{figure}[t]
  \resizebox{\hsize}{!}{
  \rotatebox{00}{\includegraphics{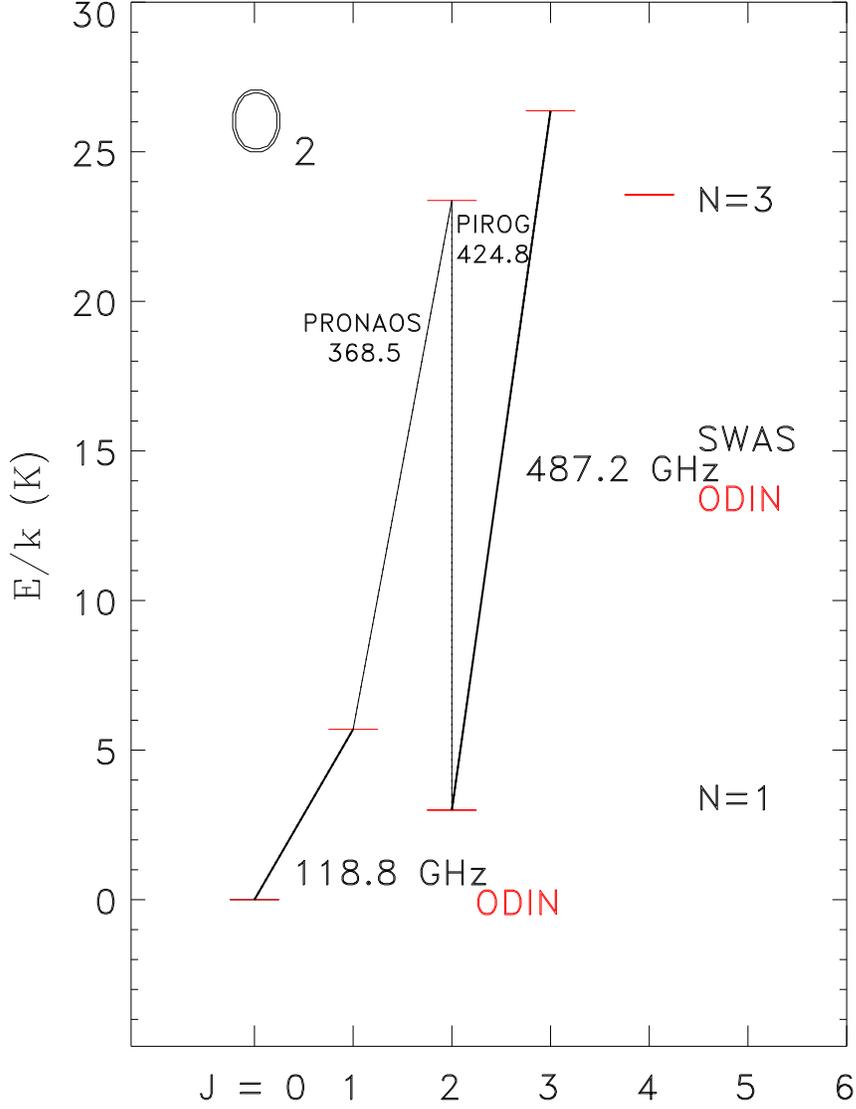}}
                        }
  \caption{Rotational energy level diagram of \molo\ for upper energies below 30\,K. Transitions which are accessible to various missions discussed in the text are also indicated (adopted from Liseau 2001).}
  \label{levels}
\end{figure}

Prior to the launch of Odin in February 2001, it was already clear from the observations by the Submillimeter Wave Astronomy Satellite (SWAS) that \water\ was significantly less abundant in the ISM than was previously thought. Even more intriguing, perhaps, was the apparently total absence of \molo\ (Goldsmith et al. 2000). However, a couple of years later, the tentative detection of the 487\,GHz line (see Fig.\,\ref{levels}) toward the dense molecular core \roa\ was announced by Goldsmith et al. (2002).

The SWAS result was challenged by Pagani et al. (2003), who were unable to confirm the presence of oxygen in \roa\ based on observations with Odin in the ground state transition of \molo\ at 119\,GHz ( Fig.\,\ref{levels}). Taking into account the fact that this involved a different line at considerably lower angular resolution toward a somewhat different position, rendered the status of the SWAS result inconclusive. 
    
Besides the negative result for the dense cloud \roa, Pagani et al. (2003) reported also on unsuccessful observations of nearly a dozen galactic sources in the \otwofull\ line at 119\,GHz. Furthermore, Wilson et al. (2205) recently published an upper limit in this line toward an extragalactic object, viz. the Small Magellanic Cloud (SMC). Contrasting with these negative results, we report here the detection with Odin, at the $5\,\sigma$ level, of the 119\,GHz line of \molo\ toward the dense cloud \roa.

\section{Odin O$_2$ Observations}

The Odin observations of \roa\ were performed during 2002 and 2003 and the total on-source integration time was 78 hours. At 119\,GHz, the half-power-beam-width of the 1.1\,m telescope is 10\amin. As back-end, a digital autocorrelator of 100\,MHz bandwidth and 125\,kHz resolution was used. In Dopler-velocity units, this corresponds to 250\,\kms\ and 0.3\,\kms, respectively. More exhaustive details regarding these observations and the subsequent complex data reductions are provided by Larsson et al. (2005).

\begin{figure}[t]
  \resizebox{\hsize}{!}{
  \rotatebox{00}{\includegraphics{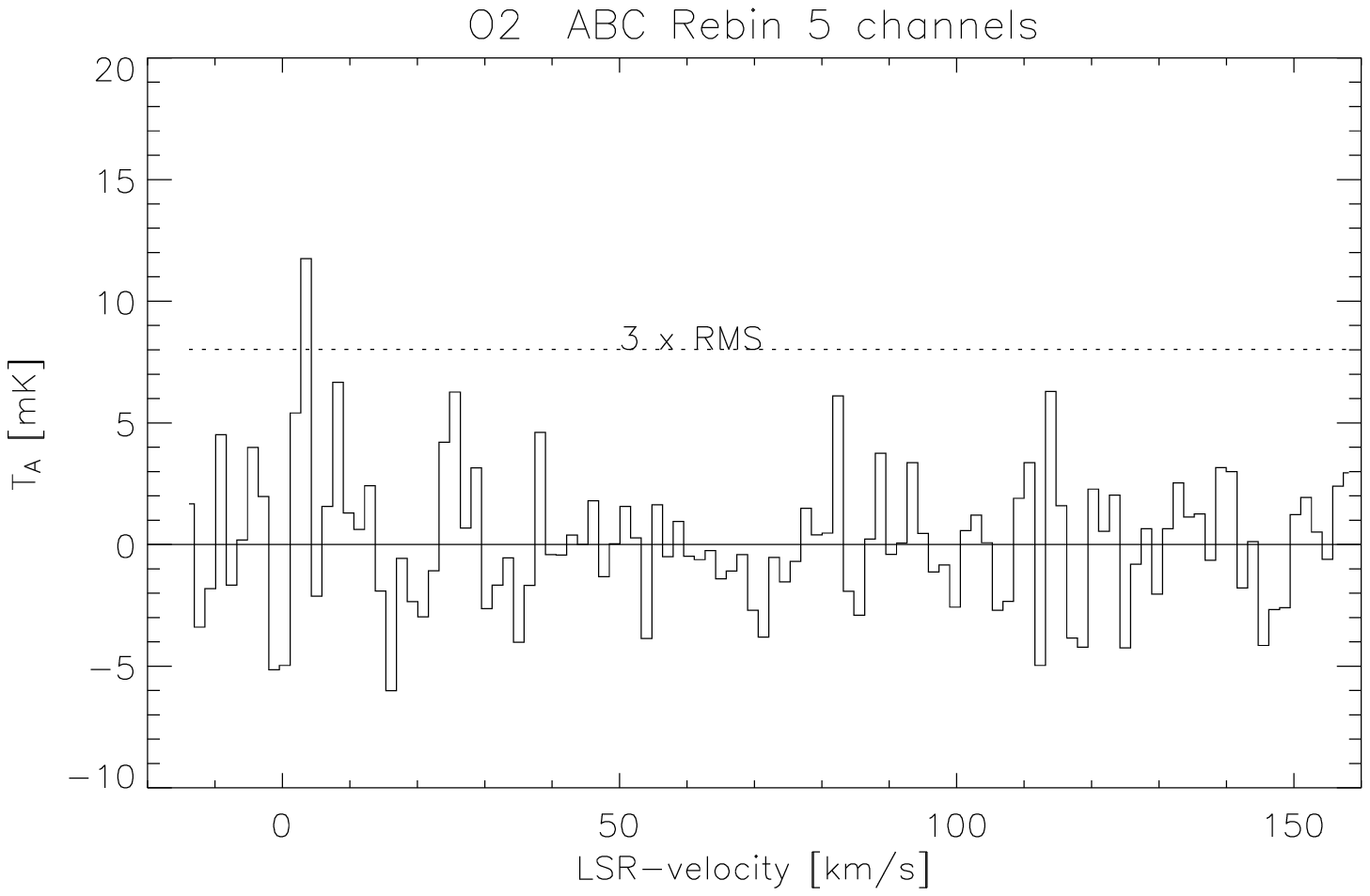}}
                        }
  \caption{The 5 channels of the \molo\ feature have been binned into 1 channel, corresponding to 
  the integrated line intensity $\int\!T_{\rm A}\,{\rm d}v$ at the $5\,\sigma$ level.
  All other features in the entire useful spectral range over 180\,\kms\ are below the $3\,\sigma$ leve, shown as the dashed horizontal line.}
  \label{o2_binned}
\end{figure}

In Fig.\,\ref{o2_binned}, showing the final stage of this data reduction, the Odin 119\,GHz spectrum is presented. At full spectral resolution, the \molo\ line occupies five autocorrelator channels, which in the figure have been binned into one channel. This is equivalent to the integrated line intensity, $\int\!T_{\rm A}\,{\rm d}v_z = (28\pm5.5)$\,mK\,\kms. 

Other Gauss-fit parameters are: peak $T_{\rm A}=(17.5\pm3.5)$\,mK, line center velocity $v_{\rm LSR}=(3.0\pm1.0)$\,\kms\ and line width $\Delta v_{\rm FWHM}=(1.53\pm0.03)$\,\kms. The velocity and width of the \molo\ line are in agreement with that of the PDR recombination line emission observed by Pankonin \& Walmsley (1978) with a \amindot{7}{8} beam, i.e. comparable to the spatial resolution of Odin. As also indicated by the C$^{18}$O observations of Frerking et al. (1989) with a \amindot{2}{3} beam, the line parameters center velocity and width of optically thin species appears constant on various angular scales. The values of both parameters are significantly different for the 487\,GHz line of Goldsmith et al. (2002).

The Odin observations at full spectral resolution, and the comparison with other melecular line data, will be presented by Larsson et al. (2005).  

\section{O$_2$ Abundance}

Of general interest is the relative concentration of various forms of oxygen in the interstellar medium and in particular that of the molecular constituents. Below, we shall examine the derivation of the molecular abundance in some detail, as widely used standard methods will encounter difficulties in this particular case, leading to erroneous results.

\subsection{Column density of oxygen}

The data for radiative ($A_{\rm ul}$) and collisional ($q_{\rm ul}$) de-excitation have been provided by Marechal et al. (1997) and by Bergman (1995 and private communication). The collisional excitation $N_J=1_0$ to $N=J$, such as $(1_0-1_1)$, is forbidden and we assume that the collisional population of the $1_1$ level occurs mainly via $3_2$ and hence, the critical density for \otwofull, viz. 

\begin{equation}
n_{\rm crit} = A_{11,~10}/q_{32,~11}(T_{\rm k}) 
\end{equation}

becomes $> 10^3$\,\cmthree\ but $< 10^4$\,\cmthree\ for $T_{\rm k} < 100$\,K and is likely lower than densities in the cloud. Local Thermodynamic Equilibrium (LTE) provides thus a reasonable assumption for the population of the lower levels. The weak \otwofull\ line is most likely optically thin and can be expected to exhibit a nearly Gaussian shape, reflecting small scale stochastic motions in the cloud. 

The observed integrated intensity of the optically thin 119\,GHz line, formed in LTE, is then given by

\begin{eqnarray}
I^{\rm obs}_{119} &  =  & \int\!T_{\rm A}\,{\rm d}v_{\rm z} \approx f_{\rm b}\,\eta_{\rm mb}\int \!T_{\rm R}\,{\rm d}v_{\rm z} \nonumber \\
                  &  =  & f_{\rm b}\,\eta_{\rm mb} \left ( \frac {h c}{2 \pi^{1/3} k} \right )^{\!3} \frac {A_{11,\,10}}{T_{11,\,10}^2} 
                          \left [ 1 - \frac {J_{\nu}(T_{\rm bg})}{J_{\nu}(T_{\rm k})} \right ]\,\frac {g_{11}\,
                          {\rm e}^{-T_{11,\,10}/T_{\rm k}}}{Q(T_{\rm k})}\,X({\rm O_2})\,N({\rm H}_2)
\end{eqnarray}

where $f_{\rm b}$ is the beam filling, $\eta_{\rm mb}$ is the main beam efficiency and \tr\ is the radiation temperature 
of the source. In LTE, the only dependence is on the temperature, which for our present purposes is well represented by a power law, i.e. for the 119\,GHz transition, the function $\Phi(T_{\rm k})$ is given by

\begin{eqnarray}
\Phi(T_{\rm k})   &  =  & \left ( \frac {h c}{2 \pi^{1/3} k} \right )^{\!3} \frac {A_{11,\,10}}{T_{11,\,10}^2} 
                          \left [ 1 - \frac {J_{\nu}(T_{\rm bg})}{J_{\nu}(T_{\rm k})} \right ]\,\frac {g_{11}\,
                          {\rm e}^{-T_{11,\,10}/T_{\rm k}}}{Q(T_{\rm k})} \nonumber \\
                  &  =  & 2.5 \times 10^{-11}\,\,T^{\,-0.67}_{\rm k}\,,\hspace*{1.0cm}T_{\rm k} > T_{11,\,10}
\end{eqnarray}

in units of K\,cm$^3$\,s$^{-1}$. In the above equations, the temperature of the back-ground radiation field is $T_{\rm bg}=2.734$\,K, the statistical weight of the upper level is $g_{11}=2J+1=3$,  the temperature of the transition is $T_{11,\,10}=5.70$\,K, the quasi-Planck function is $J_{\nu}(T)=T_{11,\,10}/[{\rm exp}(T_{11,\,10})-1]$ and the partition function is approximated by $Q(T_{\rm k})\approx kT_{\rm k}/hB$, where the rotational constant is $B=43100.460$\,MHz. 

Expressing the line intensity in K\,\kms, the column density of \molo\ is then simply found from

\begin{equation}
N({\rm O}_2)  \ge   4 \times 10^{15}\,\,T_{\rm k}^{0.67}\,\,I^{\rm obs}_{119}\hspace*{1.0cm}({\rm cm}^{-2})
\end{equation}

On the scale of the Odin beam, temperatures have variously been estimated from gas tracers (e.g., CO; Loren et al. 1983) and multi-wavelength measurements of the dust emission (e.g., Ristorcelli et al. 2005), yielding an average of about 30\,K and, hence, $N({\rm O}_2) \ge 1\times 10^{15}$\,\cmtwo. As can be seen from Eq.\,3.4, the column density of \molo\ is not very sensitive to the temperature and, e.g., an uncertainty of $\pm 10$\,K would amount to a change in $N({\rm O}_2)$ by no more than 20\%. The critical parameter for the determination of $X({\rm O}_2)$ obviously is $N({\rm H}_2)$.

\begin{figure}[t]
  \resizebox{\hsize}{!}{
  \rotatebox{00}{\includegraphics{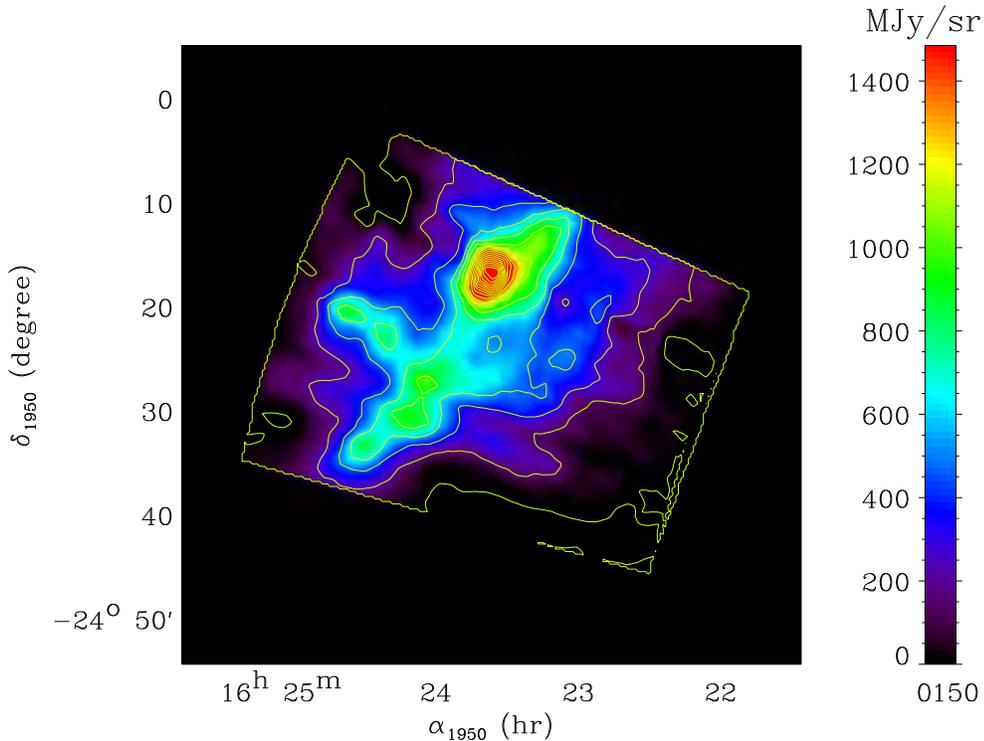}}
                        }
  \caption{Continuum observations of the \roc\ at 200\,\um\ with the 2\amin\ beam of the balloon-borne submm-telescope PRONAOS (Ristorcelli et al. 2005). \roa\ is the hot spot above the center of the figure. Intensities, $I_{\nu}$, are given in MJy\,sr$^{-1}$ (see the scale bar to the right). In addition to the 200\,\um-map shown here, full 40\amin\,$\times$\,40\amin\ maps have also been obtained at 260, 360 and 580\,\um.}
  \label{PRONAOS}
\end{figure}

\begin{figure}[t]
  \resizebox{\hsize}{!}{
  \rotatebox{00}{\includegraphics{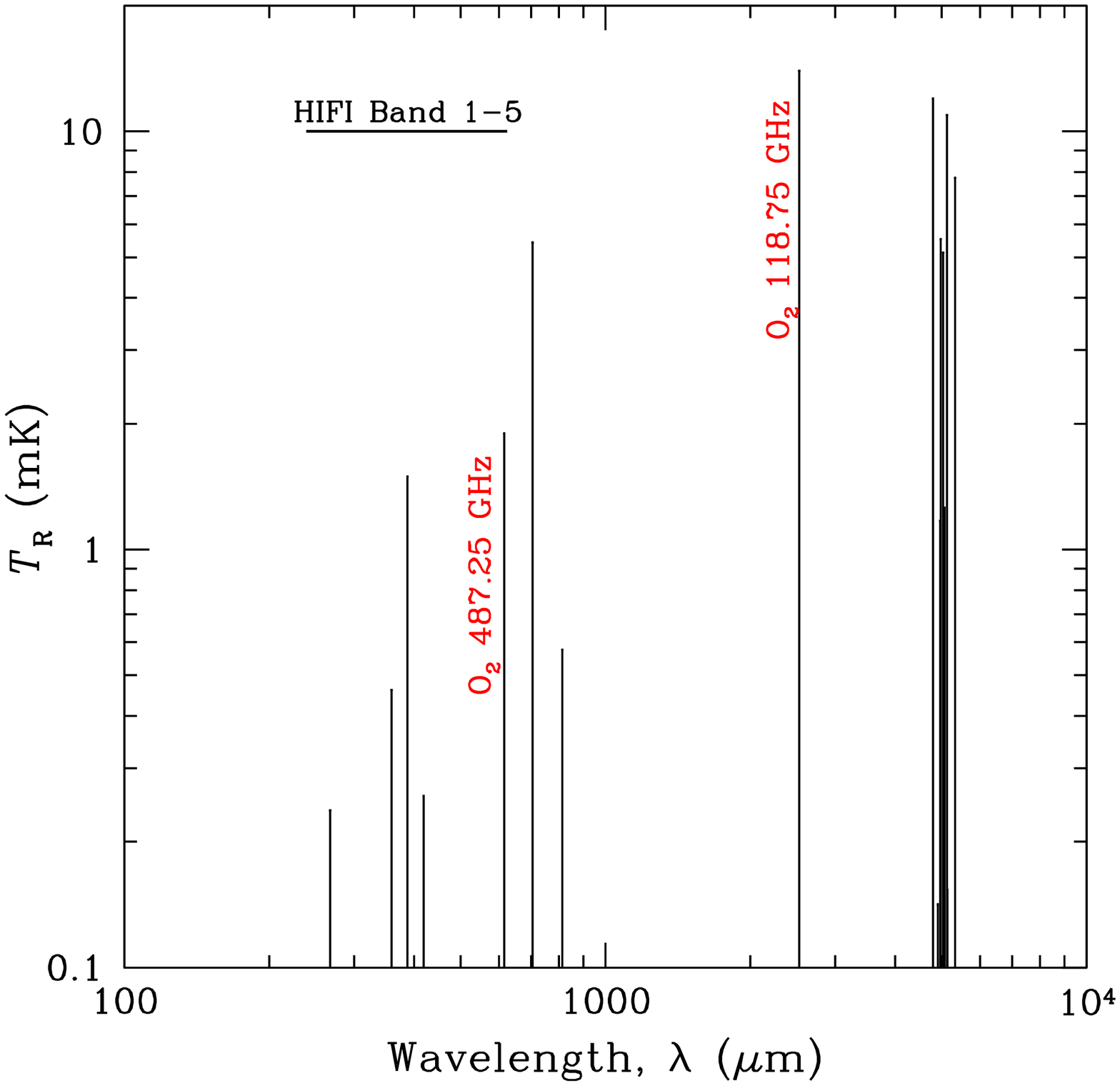}}
                        }
  \caption{Theoretical (LVG) \molo\ spectrum between 100\,\um\ and 1\,cm for \roa, based on the Odin observations in the \otwofull\ line of \roa. The uniform source is assumed to fill the 10\amin\ beam. The bar shows the frequency coverage of Herschel-HIFI, admitting the 487\,GHz ($3_3-1_2$) line at its low-frequency end (see also Fig.\,\ref{levels}), whereas the stronger 425\,GHz ($3_2-1_2$) line falls unfortunately outside the HIFI band.}
  \label{spectrum}
\end{figure}

\subsection{Column density of hydrogen}

To find the oxygen abundance, $X({\rm O}_2)=N({\rm O}_2)/N({\rm H}_2)$, we need to specify the column density of hydrogen, \molh\ (see: Eq.\,3.2). Based on the C$^{18}$O maps of Tachihara et al. (2000), Pagani et al. (2003) estimated for Odin \molo\ observations with a 10\amin\ beam the \molh-column to be $N({\rm H}_2)= 4 \times 10^{22}$\,\cmtwo. This is higher than values recently obtained from [C\,I] observations by Kulesa et al. (2005) on a much smaller, i.e. \amindot{3}{5}, scale, viz. $(2-3)\times 10^{22}$\,\cmtwo\ and those derived from submm dust continumm observations by Ristorcelli et al. (2005), viz. $1\times 10^{22}$\,\cmtwo\ for the 10\amin\ Odin beam. The earlier map of the LTE column density of C$^{18}$O by Wilking \& Lada (1983) would indicate a one order of magnitude higher $N({\rm H}_2)$. This is also the value obtained, on smaller angular scales, by Kamegai et al. (2003) from a large [C\,I] map of the \roc.
  
 Various assumptions come into these estimates, such as the abundance of C$^{18}$O or the opacity of the dust grains, leading to a considerale spread in $N({\rm H}_2)$. For the moment, we shall adopt a weighted average of $2\times 10^{22}$\,\cmtwo, a value which is associated with an uncertainty of at least a factor of two.

\subsection{O$_2$ Abundance: extended source}

Applying the estimates of temperature and \molh-column density to the \molo\ observations with Odin yield an oxygen abundance $X({\rm O}_2) \approx 5 \times 10^{-8}$, if we assume that both $\eta_{\rm mb}$ and $f_{\rm b}$ are close to unity. The corresponding rotational \molo\ spectrum between 100\,\um\ and 1\,cm is displayed in Fig.\,\ref{spectrum}. From that we infer that Odin is not capable to detect from \roa\ the other, in principle accessible, \molo\ line, i.e. the ($3_3-1_2$) line at 487\,GHz (\amindot{2}{3} beam). It is also clear that the detection of the 487\,GHz line with Herschel-HIFI\footnote{ \texttt{http://www.sron.nl/divisions/lea/hifi/} } (45\asec\ beam) would require the \molo\ source to be significantly more compact than 10\amin.

\subsection{O$_2$ Abundance: compact source}

The value of $X({\rm O}_2)$ derived above is based on an analytical model of the line excitation, where one has made the implicit assumption that the emitting region is an isothermal and homogeneous plane-parallel slab of gas. These idealized conditions would have to apply on the angular scale of 10\amin, corresponding to a physical length scale of half a parsec at the distance of the \roac. This picture is inconsistent, however, with observations on smaller scales of the gas and the dust (see, e.g., Ashby et al. 2000 and Johnstone et al. 2000, respectively), which show \roa\ to have a centrally condensed core structure. In other words, strong gradients in temperature, density and velocity field are known to be present inside the Odin \molo\ beam. In addition, UV-radiation illuminates and heats the core preferentially only from one side (Liseau et al. 1999).

Modeling the complex structure of this core is in progress. Preliminary results suggest that the line formation region of \molo\,119\,GHz is relatively compact, with a size of 22\asec. Consequently, the implied abundance is relatively high, $X({\rm O}_2)=6 \times 10^{-7}$, and the observed weakness of the line is due to heavy beam dilution (roughly $\theta^2_{\rm source}/\theta^2_{\rm beam}=10^{-3}$). Predictions of this model for the \molo\,487\,GHz line are for Odin (\amindot{2}{3} beam) a peak \ta=22\,mK and for Herschel-HIFI (45\asec\ beam) 0.2\,K. Whereas the observation of the former would imply prohibitive integration times, the latter would be a matter of minutes. We look impatiently forward to the launch of Herschel in the 2007/08 time frame.

\section{Possible implications}

The weak detection of \molo\ by Odin will likely not add to the present discussion of the energy balance of the ISM, as the role of a coolant of importance had already been altered before. For instance, the total cooling in the \molo\ lines of the spectrum of Fig.\,\ref{spectrum} amounts to a mere $7\times 10^{-5}\,L_{\odot}$.

It is conceivable, though, that this detection eventually will lead to an increased understanding of the \molo\ chemistry of the dense and cold ISM and that chemistry models may benefit from the Odin achievement. 

\section{Conclusions}

In summary, we briefly conclude the following:

\begin{itemize}
\item[i.] Very deep integrations (3.5\,mK rms) with Odin in the 1119\,GHz ground state line of \molo\ toward the dense cloud core \roa\ resulted in a $5\,\sigma$ detection of the line.  
\item[ii.] Standard analysis of an optically thin line formed in LTE leads to the beam-averaged column density of molecular oxygen for \roa, viz. $N({\rm O}_2)=1 \times 10^{15}$\,\cmtwo. 
\item[iii.] Estimates of the oxygen abundance need to rely on assumptions regarding the source size and structure with respect to the 10\amin\ beam of Odin at 119\,GHz. This results in $X({\rm O}_2) = 5 \times 10^{-8}$ for a homogeneous extended source and $X({\rm O}_2) = 6 \times 10^{-7}$ for a more realistic source structure, respectively. 
\item[iv.] This issue will most likely become resolved by observations with the future Herschel-HIFI, for which we predict an \molo\ intensity of a few tenths of a Kelvin at 487\,GHz. 
\end{itemize}

\acknowledgements The author wishes to thank the organisers for their kind invitation to this highly stimulating conference. He also wishes to express his high appreciation of the interesting and intensive discussions with many colleagues of the Odin consortium. However, Bengt Larsson deserves special mentioning, without whom the present article would not have been possible. The full list of names and affiliations of the O$_2$din Team will appear in Larsson et al. (2005).


\begin{thebibliography}{}

\bibitem[Ashby et al.(2000)]{ashby2000} Ashby M.L.N., Bergin E.A., Plume R., et al., 2000, ApJ, 539, L\,119
\bibitem[Bergman(1995)]{bergman1995} Bergman P., 1995, ApJ, 445, L\,167      
\bibitem[Frisk et al.(2003)]{frisk2003} Frisk U., Hagstr\"om M., Ala-Laurinaho J., et al., 2003, A\&A, 402, L\,27
\bibitem[Frerking et al.(1989)]{frerking1989} Frerking M.A., Keene J., Blake G.A. \& Phillips T.G., 1989, ApJ 344, 311
\bibitem[Goldsmith et al.(2000)]{goldsmith2000} Goldsmith P.F., Melnick G.J., Bergin E.A., et al., 2000, ApJ, 539, L\,123
\bibitem[Goldsmith et al.(2002)]{goldsmith2002} Goldsmith P.F., Li D., Bergin E.A., et al., 2002, ApJ, 576, 814
\bibitem[Hjalmarson et al.(2003)]{hjalmarson2003} Hjalmarson \AA., Frisk U., Olberg M., et al., 2003, A\&A, 402, L\,39
\bibitem[Johnstone et al.(2000]{johnstone2000} Johnstone D., Wilson C.D., Moriarty-Schievens G., et al., 2000, ApJ, 545, 327 
\bibitem[Kamegai et al.(2003)]{kamegai2003} Kamegai K., Ikeda M., Maezawa H., et al., 2003, ApJ, 589, 378
\bibitem[Kulesa et al.(2005)]{kulesa2005} Kulesa C.A., Hungerford A.L., Walker C.K., et al., 2005, ApJ, 625, 194
\bibitem[Larsson et al.(2005)]{larsson2005} Larsson B., et al., 2005, in preparation
\bibitem[Liseau et al.(1999)]{liseau1999} Liseau R., White G.J., Larsson B., et al., 1999, A\&A, 344, 342 
\bibitem[Liseau(2001)]{liseau2001} Liseau R., 2001, ESA-SP 460, p.\,313  
\bibitem[Loren et al.(1983)]{loren1983} Loren R.B., Sandqvist Aa. \& Wootten A., 1983, ApJ, 70, 620
\bibitem[Loren et al.(1990)]{loren1990} Loren R.B., Wootten A. \& Wilking B.A., 1990, ApJ, 365, 269
\bibitem[Mar\'echal et al.(1997)]{marechal1997} Mar\'echal P., Viala Y.P. \& Benayoun J.J., 1997, A\&A, 324, 221
\bibitem[Olberg et al.(2003)]{olberg2003} Olberg M., Frisk U., Lecacheux A., et al., 2003, A\&A, 402, L\,35
\bibitem[Pagani et al.(2003)]{pagani2003} Pagani L., Olofsson A.O.H., Bergman P., et al., 2003, A\&A, 402, L\,77
\bibitem[Pankonin \& Walmsley(1978)]{pankonin1978} Pankonin V. \& Walmsley C.M., 1978, A\&A, 64, 333
\bibitem[Ristorcelli et al.(2005)] {ristorcelli2005} Ristorcelli I., et al., 2005, in preparation
\bibitem[Tachihara et al.(2000)]{tachihara2000} Tachihara K., Mizuno A. \& Fukui Y., 2000, ApJ, 336, 519
\bibitem[Wilking \& Lada(1983)]{wilking1983} Wilking B.A. \& Lada C.J., 1983, ApJ, 274, 698
\end{thebibliography}
\end{document}